# Electrical Tuning of Neutral and Charged Excitons with 1-nm Gate

Jawaher Almutlaq*[1], Jiangtao Wang[2], Linsen Li[1], Chao Li[1], Tong Dang[2], Vladimir Bulović[2], Jing Kong[2], Dirk Englund[†1]

[1]Research Laboratory of Electronics, Massachusetts Institute of Technology,

Cambridge, 02139, MA, USA

[2]Department of Electrical Engineering and Computer Science, Massachusetts Institute of Technology, Cambridge, 02139, MA, USA

Corresponding authors: jawaher@mit.edu*; englund@mit.edu[†];

(Dated: October 30, 2023)

## Summary paragraph

Electrical control of individual spins and photons in solids is key for quantum technologies, but scaling down to small, static systems remains challenging. Here, we demonstrate nanoscale electrical tuning of neutral and charged excitons in monolayer $WSe_2$ using 1-nm carbon nanotube gates. Electrostatic simulations reveal a confinement radius below 15 nm, reaching the exciton Bohr radius limit for few-layer dielectric spacing. In situ photoluminescence spectroscopy shows gate-controlled conversion between neutral excitons, negatively charged trions, and biexcitons at 4 K. Important for quantum information processing applications, our measurements indicate gating of a local 2D electron gas in the $WSe_2$ layer, coupled to photons via trion transitions with binding energies exceeding 20 meV. The ability to deterministically tune and address quantum emitters using nanoscale gates provides a pathway towards large-scale quantum optoelectronic circuits and spin-photon interfaces for quantum networking.

# Introduction

A central challenge is achieving scalable electrical control over individual electronic quantum states for quantum technologies. Approaches like strain-induced confinement are limited to extended defects[1,2]. Band bending requires specific tip geometries[3,4] while band nesting relies on intrinsic semiconductor properties[5,6]. Gate-defined quantum dots (GDQDs) have emerged as a leading approach by enabling electrostatic isolation and control of single charges down to the single electron level [7]. Scaling efforts have arranged multiple GDQDs in linear 1D arrays with up to nine individually controllable qubits[7]. However, scaling to the large numbers (thousands to millions) of physical qubits that are needed for QIP applications from useful quantum computing or quantum networks motivates cross-bar architectures for controlling two-dimensional spin arrays.[8,9]

However, despite the demonstration of single-electron qubits in silicon-based GDQDs[10], their transport and coupling to flying qubits - to reliably transmit information between nodes over distances - remains a challenge, which is crucial for many applications, particularly quantum networks. Furthermore, practical demonstrations based on the proposed 2D architectures have been limited due to the technical challenges associated with nanofabrication. Recently, gate-defined quantum dots (QDs) have been successfully demonstrated in two-dimensional (2D) atomically thin materials [11–15], particularly in simplified geometries since the confinement is intrinsic to the semiconductor layer itself. Such materials host a plethora of various kinds of quasiparticles and some are of a particular interest for quantum information science.[16] The confinement of excitons in particular has been a point of extensive research in the field. One dimensional confinement with 50 nm gates, [13], in-plane of neutral excitons of 10 nm,[17] and quasi-zero-dimensional [18] have been demonstrated. Nevertheless, scale and fabrication with a high fill factor and individually-addressable charged emitters have not been reported yet.

Here we engineer an architecture that achieves electrical tuning in monolayers of transition metal dichalcogenides (TMDCs) at the nanoscale. This platform requires a minimal lithography and was made possible by the progress in the synthesis, transfer and alignment of carbon nanotubes (CNTs) which enabled advanced electronics including 1-nm gates [19,20] and high-density molybdenum disulfide ($MoS_2$) transistors.[21] The platform achieves three key advances: (i) molecularly precise engineering of tunable electrostatic doping-induced confinement at the nanoscale, (ii) Volt-scale exciton gating that is compatible with state-of-art CMOS technology, (iii) nanometer-scale lithography-free gates, minimizing the need for complex lithography processes. By tuning the applied vertical bias, we achieved local access and tuning from neutral to charged excitons (i.e. trions) in the monolayer limit in $WSe_2$. We have also provided a theoretical model to correlate the spatial doping density to the two-dimensional material properties at 4 K. Potential applications include deterministic on-demand single photon emitters for quantum information processing, controlled spin–photon interfaces and local probe for artificial lattices and quantum dots.

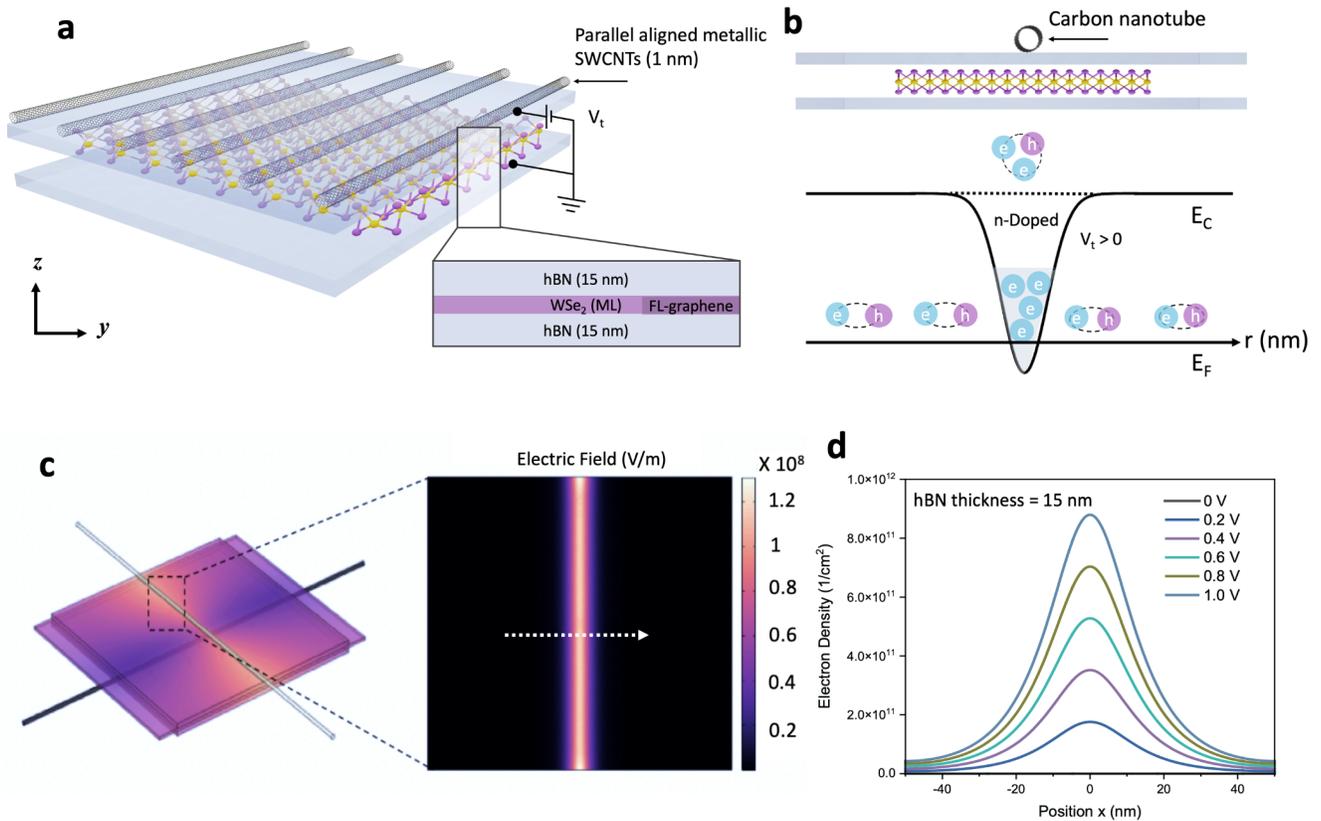

**Figure 1. Hybrid platform for local electrostatic doping. a.** schematic illustration of the device design consisting of top aligned carbon nanotubes for gating a monolayer of WSe$_2$ that is encapsulated by hBN. The WSe$_2$ is electrically grounded through a few-layer graphene flake and the bias is applied independently through the top CNTs ($V_t$), essentially forming a one-dimensional (1D) electrostatic doping profile. **b.** the relative population of the conduction band ($E_C$) across the CNT creating a local n-type when the $V_t > 0$. Due to the increased density of electrons, the excitons are converted to negative trions ($X^-$). While the schematic shows a three-particle system, depending on the nature of the 2D material, other more complex quasi particles might form including a doubly charged trion ($X^{--}$) and charged biexciton ($XX^-$). **c.** Simulated three-dimensional electrostatic potential distribution across the layers induced by the applied bias along the CNTs for $V_t = 1V$. The electrostatic potential is a function of the dielectric properties of the layers but is independent of the charge density of the active materials. Right: The closeup image shows the electric field norm along the CNT. **d.** The electron density along the dashed line in (c) when substituting for the density of states (DOS) of WSe$_2$ as a function of $V_t$ at a fixed hBN thickness of 15 nm. The charge density scales linearly with increasing $V_t$, while the confinement radius - defined as a half width at half maximum (HWHM) of the charge density peak - remains constant and is a function of the dielectric layer properties and thickness.

In this architecture, the single-walled metallic CNTs serve as nanoelectrodes to gate and control neutral and charged excitons in WSe$_2$ **(Figure 1a)**. One-dimensional CNTs are cylindrical nanostructures composed of rolled graphene sheets. The CNTs have exquisite electrical and mechanical properties including high current density, ballistic transport, and mechanical resilience making them ideal nanometer electrical contacts.[21,22] Further, they have been integrated with micro/nano systems for applications spanning areas from biosensing to high performance digital electronics.[23]

## Doping with nanometer electrodes

Here we present one-dimensional (1D) electrostatic doping based on metallic CNTs nanoelectrodes. It is worth highlighting that the electrostatic doping - when the semiconducting layer is electrically grounded - allows for sweeping the Fermi energy from p-type to n-type depending on the accumulated charges. This is in contrast with the quantum confined Stark effect (QCSE) where the perpendicular electric field across the floating semiconducting layer isolates a small region of the charge carriers and quasiparticles leading to the reduction of the corresponding density of states (DOS).[24]

Consider an electrode - a metallic SWCNT here - that is capacitively coupled to a 2D semiconductor material. By adjusting the voltage applied to the gate electrodes, the doping level can be controlled, allowing precise and local tuning of the electronic properties of the 2D material at 4 K. We assume that there is a charge distribution very close to the WSe$_2$ surface caused by the band bending perpendicular to the surface as we showed in the Methods. We solved the induced doping density numerically using the COMSOL Multiphysics finite element package. Our electrostatic simulations of the platform with metallic CNTs (radius 1 nm) and monolayer WSe$_2$ are presented in **(Figure 1c-d)**. A three-dimensional simulation of the device presented in **Figure 1c** reveals the electrostatic potential distribution when the WSe$_2$ is grounded and $V_t = 1V$. Substituting for the DOS of WSe$_2$, the charge density scales linearly with the bias and drops exponentially as a function of the dielectric thickness (i.e. hBN here), where more than 50% of the charge density drops when the thickness exceeds 2 nm (**Figure S3**). Here we define the doping parameter by taking the half width at half maximum (HWHM) of the charge density as a function of the spatial dimension perpendicular to the CNT and we use this value as a radius of doping projected on the WSe$_2$ flake. When the dielectric thickness is fixed, the radius also remains constant as a function of $V_t$, but the electron density increases linearly **(Figure 1d)**. For our platform with 15-nm thick hBN, we infer a confinement radius of less than 15 nm and a doping density around $1.8 \times 10^{12}$ 1/cm$^2$ **(Figure S4).** Our results show that the doping radius can be in the order of the exciton and trion Bohr radii at the limit of 1 nm thickness (~ 3-layer hBN) **(Figure S3,S4)**. [25,26]

Sample fabrication - The device comprises the following layers (from bottom to top): SiO$_2$ (285 nm)/Si, bottom aligned CNTs, hBN, TMD flake, few-layer graphene, hBN, and top CNTs. To fabricate the device, parallel metallic single-walled carbon nanotubes (SWNTs) (diameter ~ 1 nm) grown through chemical vapor deposition (CVD) are transferred onto pre-patterned electrodes on SiO2/Si wafer to form the bottom layer **(Figure 2b,c)**. Subsequently, a 2D monolayer flake is transferred as the active layer, encapsulated by layers of dielectric materials such as hexagonal boron nitride (hBN) **(Figure 2d)**. In this device the strain induced by the SWCNT is negligible The top layer is another bundle of parallel CNTs aligned at an angle with respect to the bottom CNTs. We selected WSe$_2$ as the semiconducting layer for this platform due to the rich spectral features that allows for better observation of a variety of charge carriers and quasiparticles.

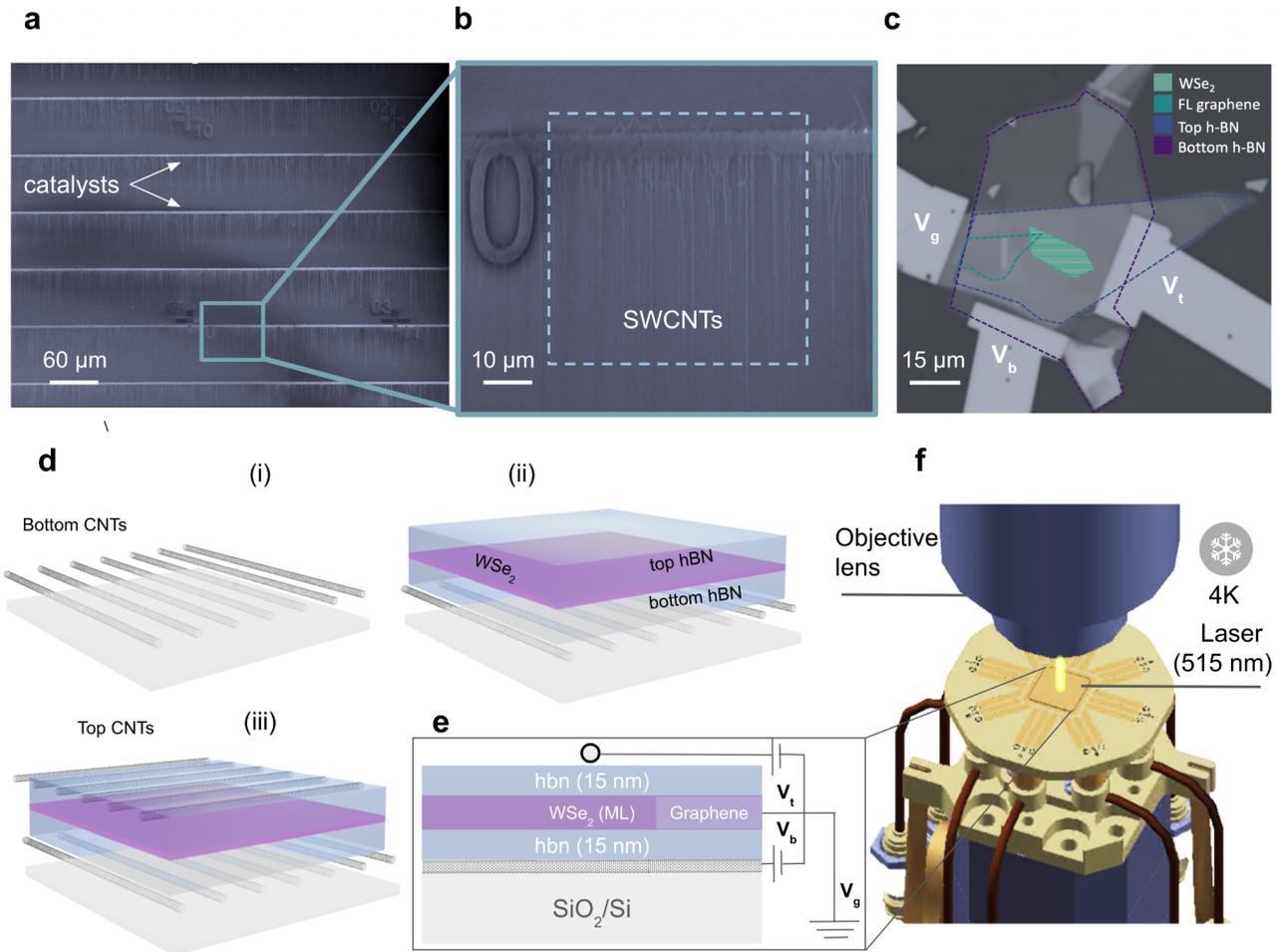

**Figure 2. Device fabrication workflow. a.** Parallel aligned SWCNTs are transferred on a SiO$_2$/Si substrate. The bright horizontal lines are the catalysts used for the growth. **b.** A closeup image of the CNTs prior to transferring the 2D layers. **c.** an optical micrograph of the device with the gate electrodes. The different layers and electrodes are marked accordingly. **d.** schematic of the fabrication process starting from the bottom CNTs in (**a**) to the final device in (**c**). **e.** A side view of the device configuration at the intersection point of the top and bottom CNTs. **f.** optical cryostation setup with electrical pads for in-situ electrical tuning. The setup is equipped with a 100x objective lens and a 515 nm excitation laser.

The spatial doping profiles of the flakes have been captured with the APD photodetector in a gate-dependent scanning confocal photoluminescence (PL) setup (the optical setup is illustrated in **Figure S1**) in which the sample was mounted in a cryostation at 4 K **(Figure 2f)**. The integration range covers all PL spectral features from 710 nm to 750 nm and we showed a represeititive PL spectrum of the flakes in **Figuere S5**. The different configurations of the applied bias are presented in **Figure 3**. We compared the profiles to SEM images to confirm the location of the CNTs, and we used the brightness of the CNTs to identify the corresponding metallicity. The bright (dark) CNTs in the SEM in **Figure 3f** are assigned to metallic (semiconducting) tubes. This is evident when comparing the AFM scans **(Figure 3d)** with the SEM images, as a few bottom CNTs were visible in the AFM but dark in the SEM. This difference in metallicity has manifested in fine or invisible PL lines **(Figure 3b-d)** indicating a lower carrier density compared to the metallic counterpart. Bright PL spots that are present in the flake in **Figure 3a-d** correspond to the same spots in the monolayer WSe$_2$ in **Figure 3e-f** with $V_t$=0 are due to bubbles and/or defects generated during the exfoliation and transfer process.

## Signature of deterministic electrostatic doping

To reveal the local doping effect on the photophysics of the charge carriers, we monitored the PL spectral of three different quasiparticles, neutral excitons (X$^o$), charged trions (X$^-$), and charged biexcitons (XX$^-$) with a fixed power of 6 uW and excitation wavelength of 515 nm at 4 K as demonstrated in **Figure S1**. We fitted and deconvoluted the peaks with Voigt functions and the results are presented in **Figure 4**. A clear spectral difference was confirmed between the points on and off the CNTs at 1V, as presented in the PL maps in **Figure 4b,c.** However, the PL spectra in **Figure S5** shows that all points on the flakes were similar when $V_t$= 0 with a small variation depending on the local environment. We found our measurements and peaks assignment to be consistent with the gate-tunable results in the literature [27], with the spectra dominated by charged quasiparticles. [28] However, all the tuned intensity and shift values presented here are almost an order of magnitude lower than the reported values in the literature. Such difference is expected due to the small gate size compared to the diffraction limit of the laser as can be inferred from the numerical analysis. Nonetheless, the small electrodes at this size limit are sufficient to induce observable spectral changes, consistent with our simulations presented earlier.

The drop in the exciton (X$^o$) PL intensity with increasing $V_t$ from 0 V to 1 V (**Figure 4d**) is accompanied by the emergence and monotonic rise of the (X$^-$) in **Figure 4e** and (XX$^-$) in **Figure 4f**, indicating an exciton-trion conversion process. We presented the results for a point off the CNTs for comparison in **Figure S6**. We rule out the possibility of exciton dissociation due to the high exciton binding energy making them robust at high applied biases. However, the I-V curve (**Figure. S7**) shows a leakage current onset around 1.2 V leading to a PL quenching potentially from enhanced non-radiative recombination owing to Auger process, Coulomb screening and exciton dissociation and scattering with the injected charge carriers.[24,29] However, the drop in the PL intensity occurs faster than the electric field onset which suggests trion dissociation events taking place at relatively high doping levels prior to the current onset. The PL intensity of the X$^-$ peak rapidly decreases beyond 1V, in contrast to the emission from the XX$^-$ that showed a small decline. This can be understood from the larger binding energy of the latter (~ 47 meV) compared to the trion peak (~ 28 meV) below the neutral exciton X$^o$ peak.

The linear dependence of the separation between X$^o$ and X$^-$ on the $V_t$ confirms the electrostatic doping in our platform and corresponds to the Fermi energy shift [30] as presented in (**Figure 4.g**). According to the many-body physics, the blue shift of the Loerntizian exciton peak is attributed to the combination of the local bandgap renormalization from the excess doping along with the reduction of the exciton binding energy.[24] The small magnitude can be understood considering the in-plane orientation of the dipole consisting of previous reports with similar device layers. The X$^-$ peak, on the other hand, redshifts with increased doping due to the Pauli blocking raising the energy barrier required to fill more electrons in the conduction band. The results are consistent with the reports on the electrostatic doping of WSe$_2$ and MoS$_2$ [31–33]

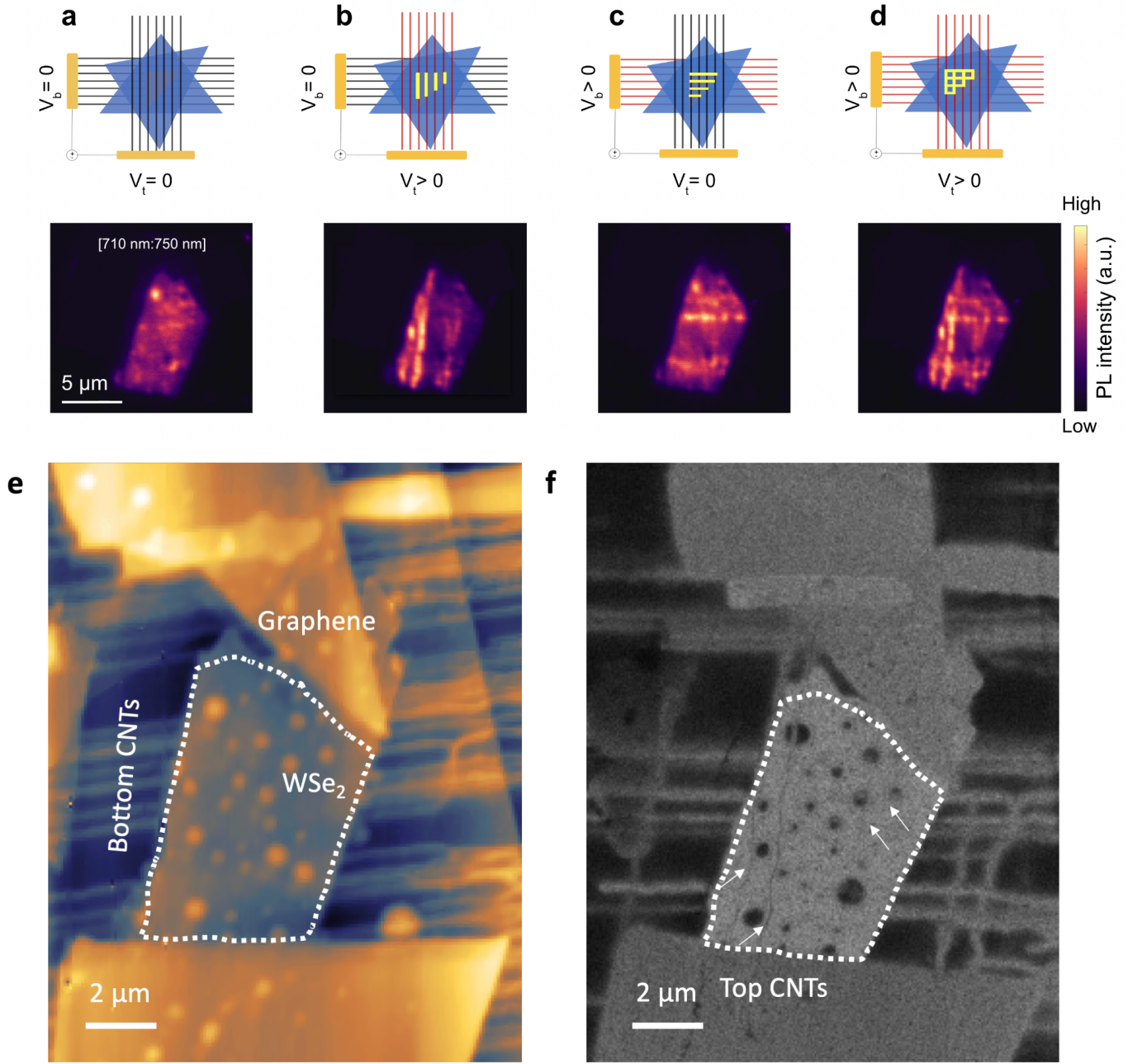

**Figure 3. Spatially resolved photoluminescence mapping from electrostatic doping with CNT electrodes. (a-d)** Confocal photoluminescence (PL) spectral mapping with integrated PL from 710 nm to 750 nm as a function of $V_t$ and $V_b$ at 4 K. **a.** $V_t = V_b = 0$ with the bright emitters from the strain induced emission generated during the exfoliation and transfer process. **b.** vertical bright lines when $V_t > 0$, $V_b = 0$. The calculated doping width is ~ 30 nm. **c.** horizontal bright lines when $V_t = 0$, $V_b > 0$, and **d.** with $V_t$, $V_b > 0$. The location of the modulated quasiparticles are compared to **e.** atomic force microscopy (AFM) topography and **f.** SEM scans to confirm the location of the top CNTs. The dotted area encircles the optically active WSe$_2$ flake and the arrows point towards the individual CNTs. Some bottom CNTs appear in the AFM but not the SEM due to the difference in metallicity.

We then investigated the assignment of the quasi particles through the dependence of the integrated PL intensity on the excitation power at $V_t$= 0 V (**Figure 4 h-i**). We have segmented the power range into low power (1-5 uW) and high power (6-20 uW) regions and then we fitted the data using the power-law $I_{PL} \propto P^{\alpha}$, where $I_{PL}$ is the integrated intensity, $P$ is the pump power, and $\alpha$ is the exponent factor. When $V_t$=0 and at low pump power, only $X^o$ shows a distinguishable peak, and the exponent is slightly above 1. From the fit, we extracted the exponent value of the high (low) power as $\alpha_{Xo} = 1.06\,(1.14), \alpha_{X-} = 1.32$, and $\alpha_{XX-} = 1.5$ (**Figure 4h**) .[34] At higher pump power, the neutral $\alpha_{Xo}$ reaches linearity and we observed a rise of the charged $X^-$ and $XX^-$. Together we explain the trend by the exciton-trion conversion from the high density excitons and the intrinsic charge doping in $WSe_2$. When then performed the same experiment at 1 V and we extracted $\alpha_{Xo} = 1.17\,(1.23), \alpha_{X-} = 1.11\,(1.37)$, and $\alpha_{XX-} = 1.79\,(1.72)$ (**Figure 4i**). The linear dependance of $\alpha_{Xo}$ and $\alpha_{X-}$ is expected from first-order radiative recombination, [35] while the superlinear growth of $\alpha_{XX-}$ arises from the combination of the linear rate of the two substituent excitons. The deviation from 2 is potentially due to the competition for the charge carrier capture from other quasiparticles. [27]

**Discussion**

In summary, we have demonstrated a scalable platform for nanoscale electrical control in 2D semiconducting materials. Using aligned CNT gates, we achieved tunable electrostatic doping to generate charged quasiparticles in $WSe_2$ at 4 K that can potentially serve as on-demand quantum emitters. Unlike strain-induced emitters, these electrically tunable emitters offer defined spectral positions and compatibility with spin qubits and spin-photon interfaces. Our theoretical calculations and experiments show confinement radii below 15 nm, approaching the exciton Bohr radius with thin hBN encapsulation. Independent tuning of the bias and dielectric thickness provides multiple knobs to control the doping profile. While current resolution is limited by diffraction, combining this platform with super-resolution imaging and resonant excitation can further improve precision. The demonstrated capability for electrical control of quantum states is compatible with various 2D materials and colloidal quantum dots. Exciting future directions include exploring quantum dots at CNT intersections and incorporating heterostructures with enhanced properties. Realizing scalable arrays of indistinguishable and electrically tunable quantum emitters will open doors to diverse quantum photonic technologies, from quantum networks to sensing and simulation.[36]

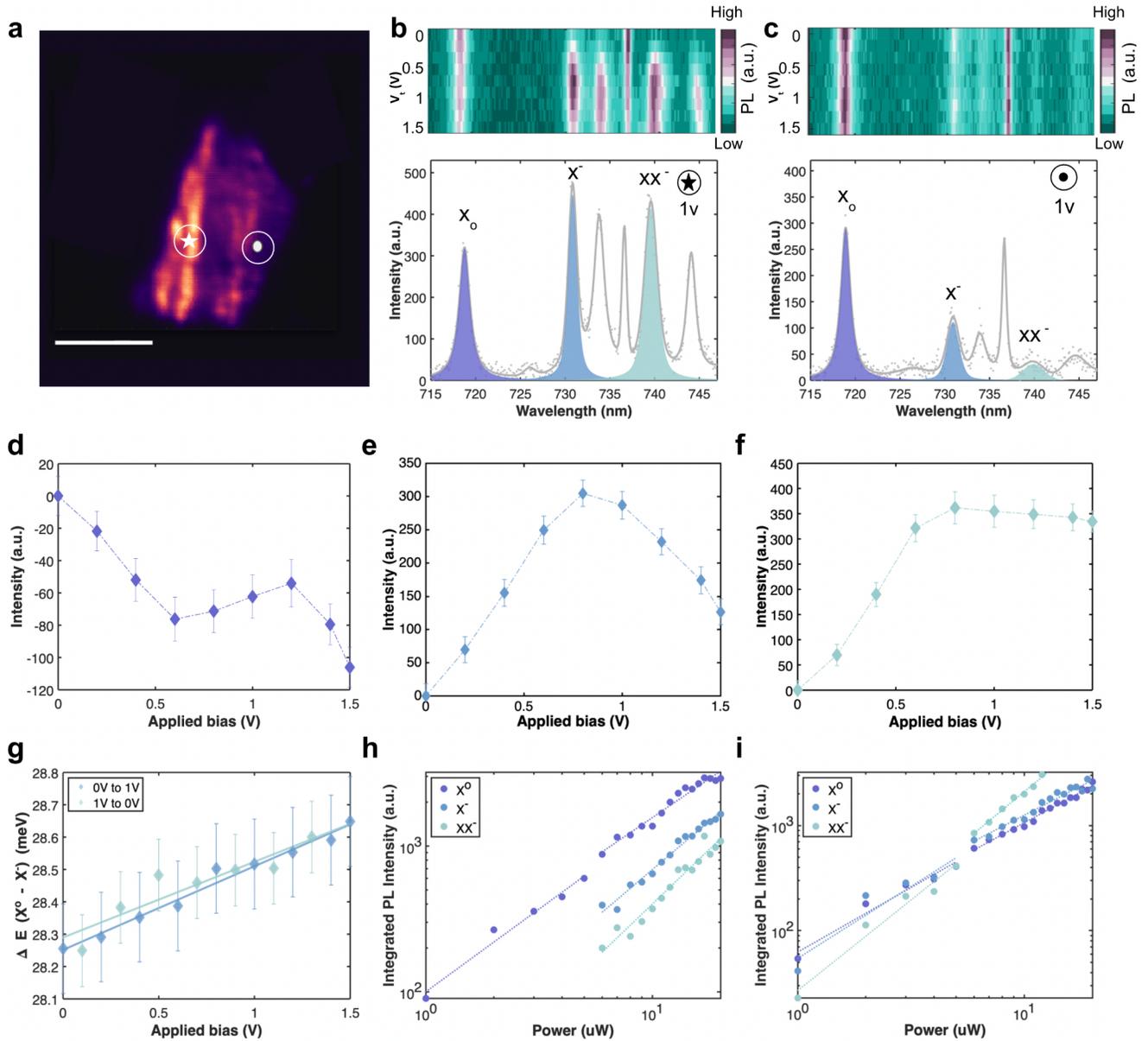

**Figure 4. Dynamic control of photoluminescence from local tunable doping with metallic carbon nanotubes. a.** reference confocal map for $V_t = 1V$ at 4 K. The star (circle) refers to the point used to measure the spectra on the CNTs (off the CNTs). Scale bar is 5 μm. **b.** a two-dimensional (2D) PL map as a function of $V_t$ revealing PL modulation of all quasiparticles at the star point on the map with $V_t > 0$, and PL spectrum point when $V_t = 1V$ **c.** 2D map from the circle point away from the CNTs with the absence of any peak modulation when $V_t > 0$, and PL spectrum point when $V_t = 1V$. **d.** PL intensity of neutral exciton ($X^o$) subtracted from the peak when $V_t = 0V$ to monitor the PL modulation shows a slight reduction as a function of $V_t$. In parallel, there is an increase in the PL intensity of **e.** ($X^-$) and **f.** ($XX^-$). We attributed the reduction in the ($X^o$) to the exciton-trion conversion process. However, the magnitude is different because the ($X^o$) peak is averaged by all points away from the CNT but within the laser excitation diffraction limit, unlike the charged excitons that only appear along the CNTs. **g.** The energy

shift between the (X$^o$) and (X$^-$) linearly increases as a function of V$_t$ due to the redshift of (X$^-$) PL peak, further confirming the electrostatic doping. **h.** Double logarithmic plot of PL intensity as a function of excitation power with excitation wavelength = 515 nm at the star point for V$_t$ = 0V. The neutral exciton shows a clear peak with linear dependence at relatively higher power ($\alpha_{X_o} = 1.06$) **i.** At V$_t$ = 1V, the charged excitons dominate at higher powers with super linear dependence while the neutral exciton remains almost linear.

## Acknowledgment

This work was supported by the Army Research Office MURI (Ab-Initio Solid-State Quantum Materials) Grant no. W911NF-18-1-043. The device fabrication and characterization were carried out in part through the use of MIT.nano's facilities. We thank Hamza Raniwala for the useful discussion about multiphysics simulation and Ian Christen for his help during the optical measurements. J.A. acknowledges the fund from KACST-MIT Ibn Khaldun Fellowship for Saudi Arabian Women at MIT, and from Ibn Rushd Postdoctoral award from King Abdullah University of Science and Technology (KAUST).

2020;20: 1869–1875.

## Methods

**Growth of single-walled carbon nanotube (SWCNT) array:** To grow SWCNT arrays, we used a patterned ST-cut single-crystalline quartz substrate that was annealed at 1000°C in an oxygen atmosphere for 10 hours. The substrate was prepared using a lithography process, and 0.4 nm Fe catalyst strips were deposited using e-beam evaporation. The growth was conducted in a 2-inch sliding furnace (OTF-1200X-80SL, MTI corp.). Before the growth, the furnace was purged with 1000 sccm of Ar for 12 minutes, and then rapidly slid over the sample to increase the growth temperature to 800°C within 8 minutes. After stabilizing for 5 minutes, a feeding gas mixture of 360 sccm Ar, 40 sccm $H_2$, 200 sccm $CH_4$, and 0.1 sccm $C_2H_4$ was introduced. However, the supersaturation of this feedstock was not sufficient to initiate SWCNT growth. Therefore, after 8 minutes, the total flow rate was suddenly reduced to 6 sccm, increasing the heating time of the feedstock and resulting in further activation of the carbon precursor and higher supersaturation. This initiated the collective growth of the SWCNT array. After 2 minutes of growth, the furnace was slid away, and the sample was quickly cooled in a flow of 1000 sccm Ar. The transfer of SWCNT was done following the report from Jiao et al.[1]. However, we have modified the method using wet transfer at cold temperature, which is faster than the conventional way and prevents degradation of CNT performance.

**Preparing the CNT for device integration:** First, 400 nm of PMMA 950 A4 resist was spin coated on the substrate and baked at 180 °C for 2 minutes. An 8-nm thick Au thin film was then thermally evaporated on top of the resist as a charge dissipation layer. The sample was exposed by electron beam lithography (EBL) with an area dose of 1400 µC/cm$^2$. After EBL, the Au charge dissipation layer was removed in Au etchant for about 15 seconds. The sample was then developed in methyl isobutyl ketone (MIBK): isopropyl alcohol (IPA) 1:3 solution for 90s, followed by a rinse in IPA and nitrogen blow dry.

**Device fabrication:** bottom carbon nanotubes were transferred on a 285 nm $SiO_2$/Si substrate pre-cleaned with piranha. The heterostructure device was assembled following a dry transfer method. First, single-layer WSe2 and and single-layer and few-layer hBN and graphene were mechanically exfoliated on 300 nm SiO2/Si substrate with an adhesive tape. The transfer process was performed using a stacking station with a microscope and piezo controllers for sub-micron alignment. The layers were picked up using a clear polydimethylsiloxane (PDMS) stamp covered with a thin polycarbonate (PC) layer. Once all layers were on the stamp, the heterostructure was then released on the aligned CNTs by melting down the PC film with the proper alignment to ensure contact with the electrodes without shorting the device. The device was left overnight to avoid washing away the layers before washing off the film. The PC film was then removed in chloroform bath for 6 hours followed by a few cycles of isopropanol wash and gentle blow dry with the nitrogen gun.Two electrodes were fabricated to contact the top CNTs separatelyand the graphene/$WSe_2$ flake. Electrodes were 10 nm Cr / 90 nm Au.

**Topology characterization:** The surface and topology were characterized using ZEISS GeminiSEM and Cypher VRS AFM. Raman spectra were recorded with a micro‑Raman spectrometer (Renishaw inVia plus).

**Photoluminescence measurements and setup:** The optical measurement is conducted with a home-built confocal microscopy with a crysostation as reported in Linsen Li et at. [2] .Briefly, two-axis galvanometer mirrors scan the excitation beam from a laser source (Cobolt) 515 nm to excite the sample. An objective lens (ZEISS, NA = 0.9, 100x) enables optical access to the samples. Collected photons are either sent to an avalanche photodetector (APD) and/or measured by the spectrometer (SP-2500i, Princeton Instruments). All optical measurements were carried out at 4 K unless stated otherwise.

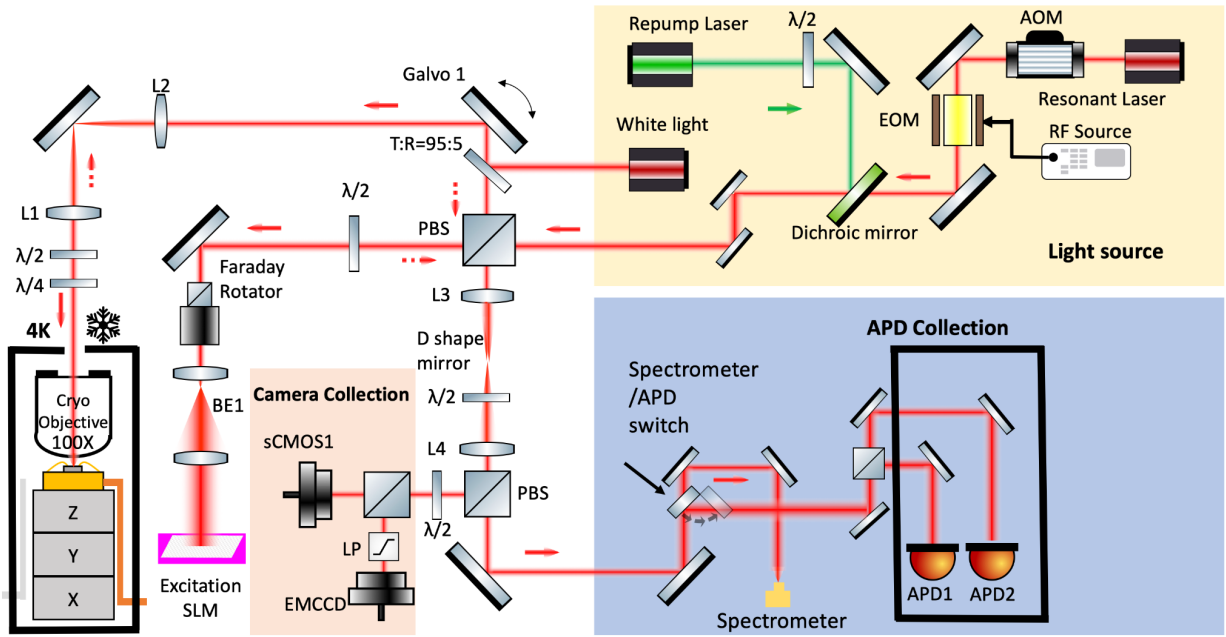

**Figure S1.** The optical measurement setup for the hybrid platform for local electrostatic doping

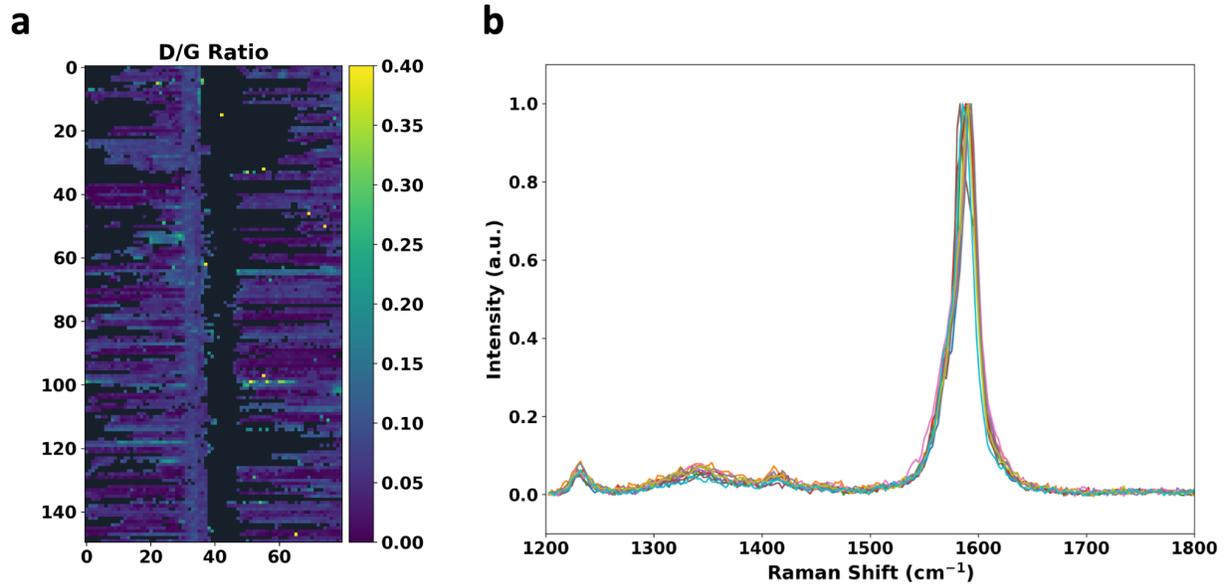

**Figure S2. a.** 2D Raman map shows a D/G ratio reaching zero highlighting the quality and low defect density in the aligned CNTs **b.** Raman of as-grown SWCNTs on quartz substrate.

**Simulation parameters**

Electrostatic simulations were carried out using the electrostatic and semiconductor modules in COMSOL multiphysics simulation. The simulation parameters are listed below. All simulations were carried out for a fixed temperature of 4 K.

**Table S1.** Parameters used for simulating the heterostructure in COMSOL

|  | WSe$_2$ (monolayer) | hBN | CNT |
|---|---|---|---|
| Thickness | 0.65 nm (monolayer) | 15 nm | 1 nm |
| Relative permittivity [3] | 7.25 (in-plane)<br>5.16 (out-of-plane) | 6.86 (in-plane)<br>3.76 (out-of-plane) | - |
| Bandgap | 1.725 eV | 5.5 eV | - |

**Electrostatic simulation of CNT-WSe2 platform**

We have applied the Thomas–Fermi approximation and we assumed a fixed temperature of 4 K. Briefly, the surface charge density is calculated as following

$$\sigma = \int_0^{+\infty} - e^2 \cdot DOS \cdot \phi_S(z) \, dZ \tag{1}$$

where $\phi_S$ is the surface potential, $DOS$ refers to the density of state (assumed to be constant with the range of surface potential) and calculated as [4]

$$DOS(E) = \frac{g_s g_v m^*}{2\pi \hbar^2} \tag{2}$$

where $g_s$ and $g_v$ are the spin and valley degeneracy, respectively. The $z$ represents the position starting from the metal surface. On the other hand, the surface charge density should be proportional to the electric field perpendicular to the surface

$$\sigma = - \varepsilon_0 \varepsilon_r E(z), \quad z = 0 \tag{3}$$

The calculated thickness of the surface charge is about 0.1 nm. Compared with the thickness of $WSe_2$ of 0.65 nm, it is evidence of strong screening effect when $V_t > 0$. By solving for the the boundary condition of the carbon nanotube and the boundary condition along the $WSe_2$, the charge density inside the $WSe_2$ film can be calculated as

$$\varepsilon_r \varepsilon_0 \frac{\partial \phi}{\partial z}\bigg|_{WSe_2} = \int_0^{+\infty} - e^2 \cdot DOS \cdot \frac{1}{e^{\frac{E-\phi}{kT}} + 1} \cdot dE \tag{4}$$

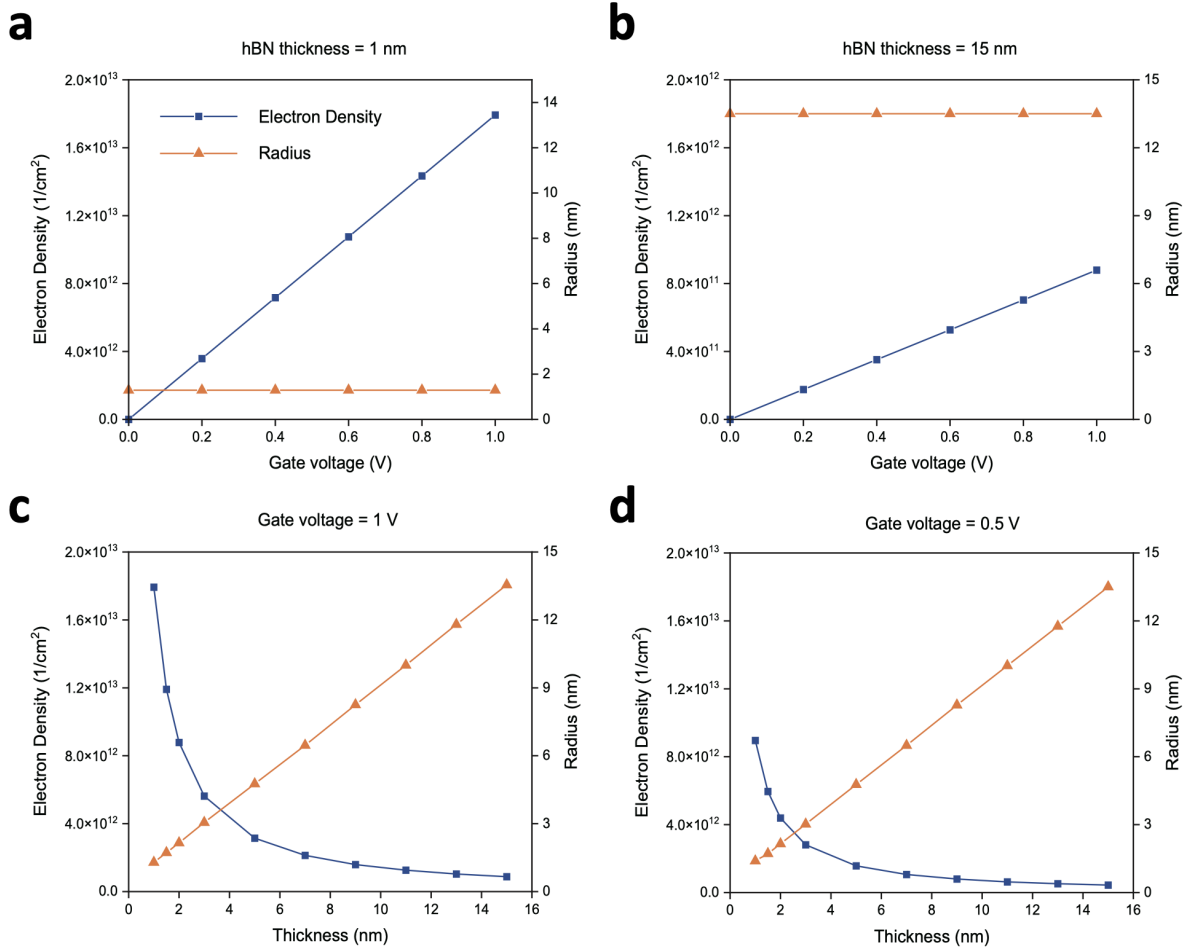

**Figure S3. Electron density and confinement radius induced by electrostatic doping. (a,b)** The electron density and confinement radius as a function of $V_t$ at a fixed thickness of 1 nm and 15 nm, respectively. **(c,d)** The electron density and confinement radius as a function of dielectric layer thickness from 1 - 15 nm at a fixed $V_t$ of 1V and 0.5 V, respectively.

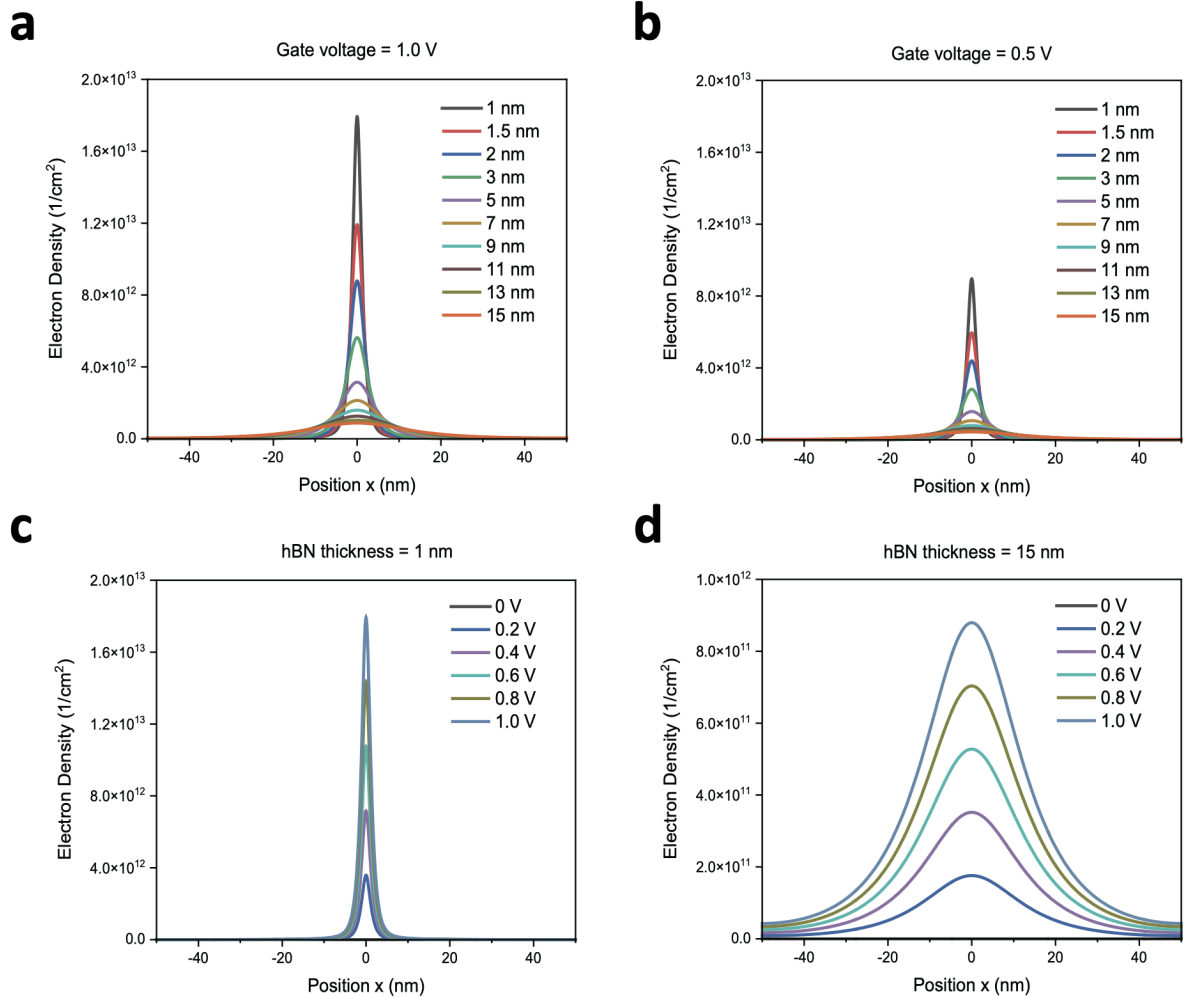

**Figure S4. Spatial distribution of the charge density and confinement electrostatic doping from the metallic CNTs.** The electron density magnitude and spatial distribution radius as a function of the position across the CNT where x = 0 is the CNT center. **(a,b)** The doping as a function of dielectric layer thickness from 1 - 15 nm at a fixed $V_t$ of 1V and 0.5 V, respectively. **(c,d)** The doping as a function of $V_t$ from 0 - 1V at a fixed thickness of 1 nm and 15 nm, respectively.

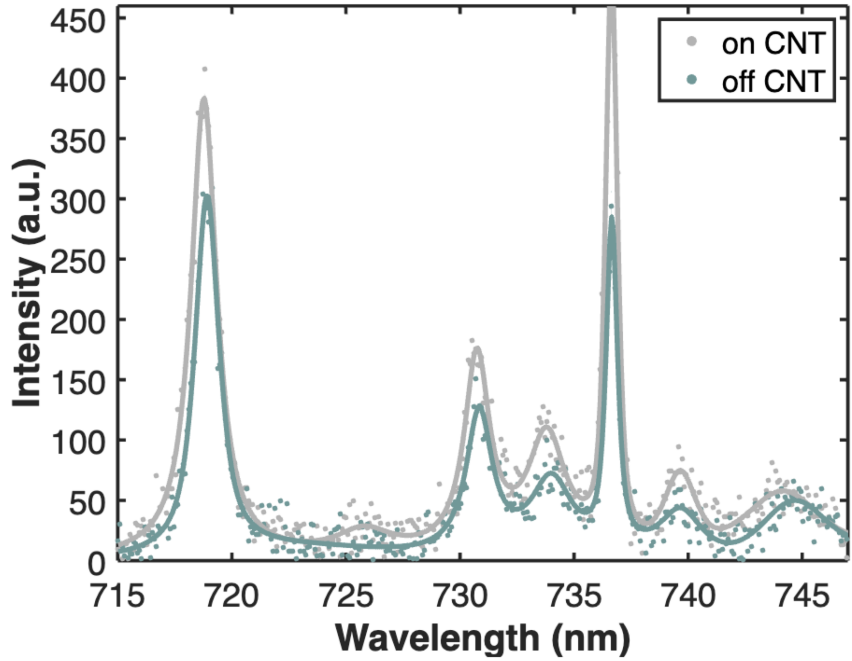

**Figure S5.** A representative PL spectrum at $V_t=0$ on the CNT (gray) and off the CNT (green) measured at 4 K.

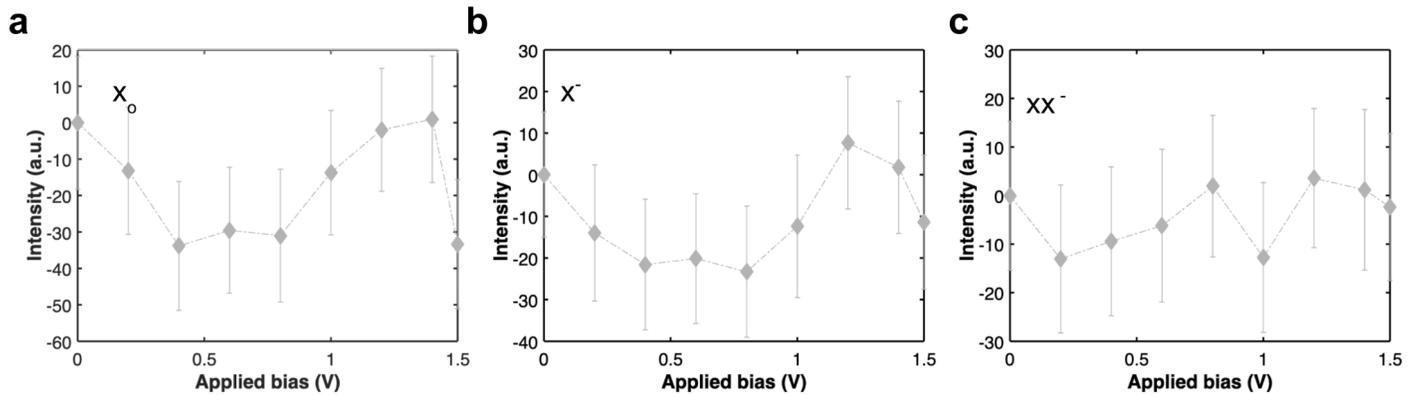

**Figure S6.** The change in the PL of the **a.** neutral exciton ($X^o$), **b.** charged exciton ($X^-$), and **c.** charged biexciton ($XX^-$) in $WSe_2$ at 1V at a point away from the CNT gates (off CNT).

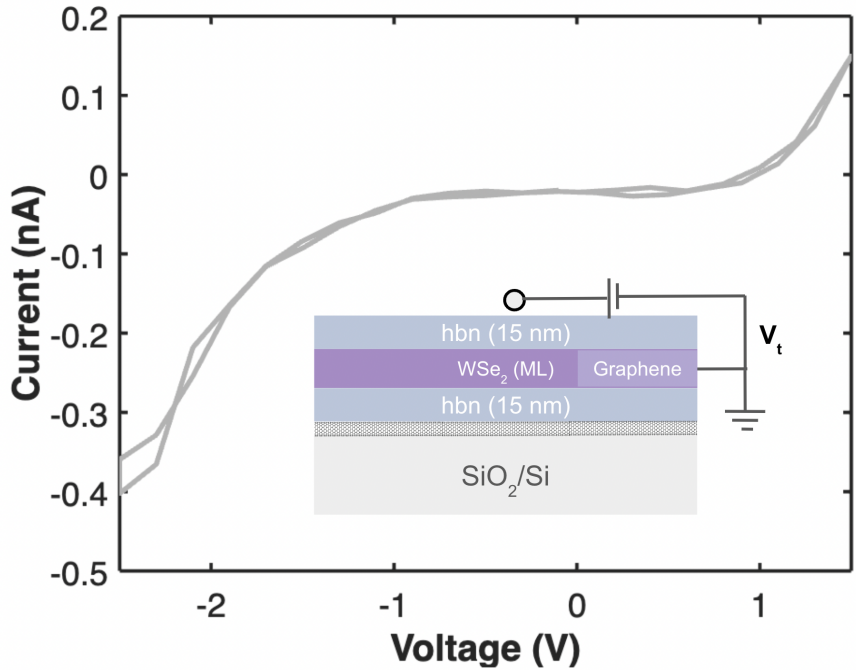

**Figure S7.** The I-V curve is measured between the top CNTs ($V_t$) and graphene/WSe$_2$.